\documentclass{article}
\usepackage{spconf,amsmath,graphicx}
\usepackage{cite}
\usepackage{amssymb,amsfonts,amsthm,bm}
\usepackage[usenames,dvipsnames]{xcolor}

\usepackage{booktabs}
\usepackage{graphicx}
\usepackage{multirow}
\usepackage{stfloats}
\usepackage{array} 
\usepackage{caption}

\definecolor{mygreen}{rgb}{0,0.5,0}
\definecolor{darkblue}{RGB}{0,0,150}
\title{A Bayesian Permutation training deep representation learning method for speech enhancement with variational autoencoder}
\name{Yang Xiang$^{ \star\dagger}$, Jesper Lisby Højvang$^{\dagger}$, Morten Højfeldt Rasmussen$^{ \dagger}$, Mads Græsbøll Christensen$^{\star}$ \thanks{This  work  was partly supported  by   Innovation  Fund Denmark (Grant No.9065-00046).}} 
\address{$^{\star}$ Audio Analysis Lab, CREATE, Aalborg University, Aalbory, Denmark  \{yaxi,mgc\}@create.aau.dk\\
	$^{\dagger}$ Capturi A/S, Aarhus, Denmark  \{yxi,jlh,mhr\}@capturi.com}
\vspace{-0.35cm}
\begin{document}
\ninept
\maketitle
\ninept
\begin{abstract}
Recently, variational autoencoder (VAE), a deep representation learning (DRL) model, has been used to perform speech enhancement (SE). However, to the best of our knowledge, current VAE-based SE methods only apply VAE to model {speech} signal, {while} noise is modeled using the traditional non-negative matrix factorization (NMF) model. One of the most important reasons for using NMF is that these VAE-based methods cannot disentangle the speech and noise latent variables from the observed signal. Based on Bayesian theory, this paper derives a novel variational lower bound for VAE, which ensures that VAE can be trained in supervision, and can disentangle speech and noise latent variables from the observed signal. This means that the proposed method can apply the VAE to model both speech and noise signals, which is totally different from the previous VAE-based SE works.  More specifically, the proposed DRL method can learn to impose speech and noise signal priors to different sets of latent variables for SE. {The experimental results show that the proposed method can not only disentangle speech and noise latent variables from the observed signal, but also obtain a higher scale-invariant signal-to-distortion ratio and speech quality score than the similar deep neural network-based (DNN) SE method.}  

\end{abstract}
%
\section{Introduction}
\label{sec:intro}
In real-world environments, speech signals are often distorted due to the presence of background noise. To reduce the effects of noise, speech enhancement (SE) techniques have been developed \cite{loizou2013speech, wang2018supervised} to improve the quality and intelligibility of {an} observed signal. 

Currently, many single-channel SE algorithms have been proposed, which include some unsupervised algorithms \cite{gerkmann2011unbiased,jensen2015noise} and supervised algorithms \cite{xiang2020nmf, kavalekalam2018online}. {However, these methods usually apply linear processes to model complex high-dimensional signal, which is not always reasonable in practical applications \cite{xu2014regression}. Thus, non-linear deep neural network (DNN) models have been developed. As shown in \cite{xu2014regression,wang2014training,wang2018supervised,luo2019conv,li2021icassp}, DNN-based methods can achieve better SE performance than traditional linear models.} However, their generalization ability is not often satisfactory for the unseen noise {conditions} \cite{wang2018supervised}. 

Recently, {deep probabilistic generative models have been investigated to improve the DNN's generalization ability for SE}, such as generative adversarial networks (GAN) \cite{xiang2020parallel} and {the} variational autoencoder (VAE) \cite{kingma2013auto,bando2018statistical}. VAE can learn the probability distribution of complex data and perform efficient approximate posterior inference, so VAE-based SE algorithms {have} been proposed \cite{leglaive2018variance,bando2018statistical,leglaive2019semi}. However, the VAE of these methods is trained in an unsupervised manner on speech only, and the noise is modeled by an NMF model {because these methods cannot disentangle the speech and noise latent variables from the observed signal. This means that these algorithms are not robust \cite{carbajal2021guided},} and their SE performance is limited compared to DNN-based supervised methods\cite{bando2018statistical}. To mitigate this problem, supervised VAE-based SE methods {have} been proposed. In \cite{carbajal2021guided,fang2021variational}, a supervised classifier\cite{carbajal2021guided} and a supervised noise-aware training strategy\cite{fang2021variational} are introduced to the training of speech VAE. The purpose is to obtain a more robust speech latent variable from the observed signal. {However, the noise is still modeled by {a} linear NMF model because it is a difficult task to disentangle the speech and noise latent variables from the observed signal \cite{fang2021variational}.}

{Learning interpretable latent representation is a challenging but very useful task because it can explain how different factors
influence the speech signal, which is important in speech-related applications\cite{hsu2017learning}.} In \cite{hsu2017learning}, a latent space arithmetic operation was derived to modify the speech attributes (phonetic content and speaker identity). \cite{hsu2017unsupervised} proposed an unsupervised method to distinguish different latent variables and generate new latent variables for the ASR application. \cite{hsu2017unsupervised_learning} applied VAE to learn the sequence-dependent and sequence-independent representations. However, interpretable latent representation is rarely considered in current SE algorithms \cite{xu2014regression,wang2014training,luo2019conv,li2021icassp}.

Inspired by previous work, in this paper, we propose a Bayesian permutation training method for SE. The proposed method can disentangle the latent speech and noise variables from the observed signal in a supervised manner and conduct the mapping between latent variables and targets{, even though} this is a difficult task \cite{fang2021variational}. We hypothesize that disentangling latent variables can improve the performance of the supervised DNN-based SE method. To achieve this, a clean speech VAE (C-VAE) and a noise VAE (N-VAE) are separately pre-trained without supervision. After that, based on {Bayesian} theory and our derived variational lower bound, we use the two pre-trained VAEs to train a noisy VAE (NS-VAE) in a supervised manner. The trained NS-VAE can learn the latent representations of the speech and noise signal. When we conduct SE, the trained NS-VAE is first used to predict the latent variables of the speech and noise signal. Then, the two latent variables are independently used as the decoder input of {the} C-VAE and N-VAE to estimate the corresponding speech and noise. Finally, the enhanced signal can be acquired by {direct} speech waveform reconstruction  or with post-filtering methods. Compared to previous VAE-based SE methods \cite{leglaive2018variance,bando2018statistical,leglaive2019semi,carbajal2021guided,fang2021variational} and interpretable latent representation learning methods \cite{hsu2017learning,hsu2017unsupervised,hsu2017unsupervised_learning}, the proposed method derives a novel variational lower bound to ensure that supervision training can be used for VAE, and VAE can disentangle different latent variables to model noise for SE. The derived supervised lower bound is very different from previous VAE-based methods\cite{kingma2013auto,leglaive2018variance,bando2018statistical,leglaive2019semi,carbajal2021guided,fang2021variational, hsu2017learning,hsu2017unsupervised,hsu2017unsupervised_learning} that are trained on an unsupervised variational lower bound, which increases the robustness of different learned latent variables \cite{fang2021variational}. Moreover, our learned latent variables {are attributed} to different types of signal, so each single latent variable that is generated by NS-VAE can be {used} to generate the corresponding speech or noise {signal}, and their combination can generate noisy speech.
\vspace{-0.3cm}
\section{Problem Description}
\vspace{-0.2cm}
\label{sec:problem}

In this work, we {aim to} perform SE in an additive noisy environment. Thus, the signal model can be written as
\begin{equation}
  \scriptsize
 \setlength{\abovedisplayskip}{3pt}
  y(t) = x(t)+d(t),
  \label{time_noisy_model}
   \setlength{\belowdisplayskip}{2pt}
\end{equation}
where \(y(t)\), \(x(t)\), and \(d(t)\) represent the observed, speech, and noise signal, respectively, and $t$ is the time index. {Log-power spectrum (LPS) is suitable for direct signal estimation\cite{xu2014regression}, so we use it as a feature for SE. The LPS of \(y(t)\), \(x(t)\), and \(d(t)\) is written as $Y(f,n)$, \(X(f,n)\), and \(D(f,n)\), respectively. Here, $f \in [1, F]$ and $n \in [1, N]$ denote the frequency bins and time frame indices, respectively.} Collecting $F$ frequency bins and $N$ time frames, we can obtain the LPS dataset $\mathbf{Y}_N$, $\mathbf{X}_N$ and $\mathbf{D}_N$ with $N$ samples, where ${\mathbf{Y}}_N=[\mathbf{y}_1,\cdots, \mathbf{y}_n, \cdots, \mathbf{y}_N]$ and $\mathbf{y}_n=[Y(1,n), \cdots, Y(f,n),\cdots, Y(F,n)]^T$, $\mathbf{x}_n$ and $\mathbf{d}_n$ are defined similarly to $\mathbf{y}_n$. $\mathbf{X}_N$ and $\mathbf{D}_N$ are defined similarly to $\mathbf{Y}_N$. For simplicity, we use $\mathbf{y}$, $\mathbf{x}$, and $\mathbf{d}$ to represent a sample in dataset $\mathbf{Y}_N$, $\mathbf{X}_N$ and $\mathbf{D}_N$, respectively. 
\begin{figure}[!tbp]
  \centering
  \setlength{\abovecaptionskip}{0.1cm}
  \centerline{\includegraphics[scale=0.5]{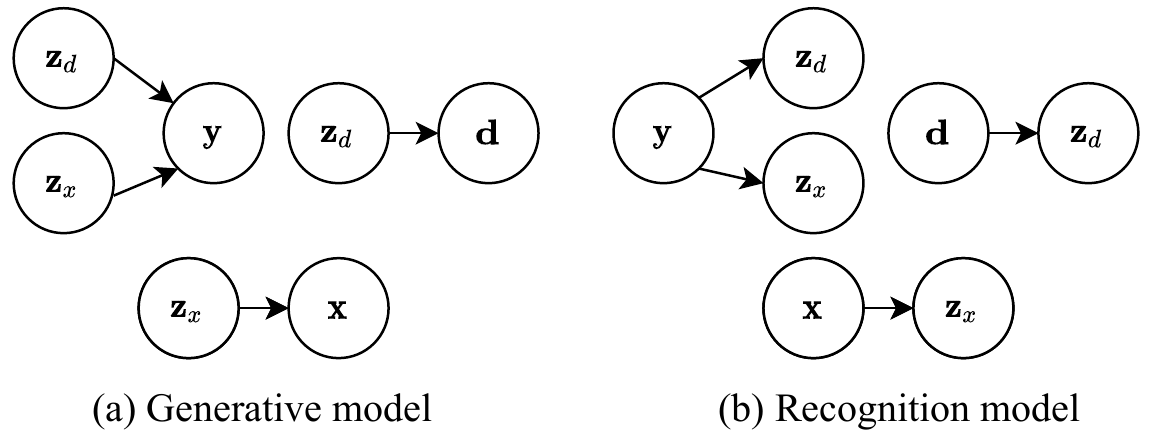}}
  \setlength{\belowcaptionskip}{3pt}
  \caption{Graphical illustration of the proposed model.}
  \label{fig:Bayesian_model}
  \vspace{-0.9cm}
\end{figure}
In the proposed VAE model,  we assume that $\mathbf{y}$ is generated from a random process involving the speech latent variables $\mathbf{z}_x$ and the noise latent variables $\mathbf{z}_d$, where the observed speech conditional prior distribution can be written as $q(\mathbf{y}|\mathbf{z}_d, \mathbf{z}_x )$. The dimensions of vectors $\mathbf{z}_x$ and $\mathbf{z}_d$ are $L_x$ and $L_d$, respectively. The dataset of $\mathbf{z}_x$ and $\mathbf{z}_d$ {is written} as $\mathbf{Z}_{xN}$ and $\mathbf{Z}_{dN}$ with $N$ samples. Here, we assume that the latent variables $\mathbf{z}_x$ and $\mathbf{z}_d$ {are} independent. Additionally, $\mathbf{x}$ and $\mathbf{d}$ are drawn from the speech prior distribution $q(\mathbf{x}|\mathbf{z}_x )$ and {the} noise prior distribution $q(\mathbf{d}|\mathbf{z}_d )$, respectively. The whole generative process is illustrated in Fig.~\ref{fig:Bayesian_model}(a). In VAE, since exact posterior inference is intractable, we {propose} that $\mathbf{z}_x$ and $\mathbf{z}_d$ can be estimated from speech and noise posterior distributions $p(\mathbf{z}_x|\mathbf{x})$ and $p(\mathbf{z}_d|\mathbf{d})$, respectively, or they can also be estimated from the noisy speech posterior distributions $p(\mathbf{z}_x|\mathbf{y} )$ and $p(\mathbf{z}_d|\mathbf{y} )$. Here, we assume $p({\mathbf{z}_x},{\mathbf{z}_d}|\mathbf{y}) = p(\mathbf{z}_x|\mathbf{y})p(\mathbf{z}_d|\mathbf{y})$, which ensures that noise can be modeled by non-linear VAE rather than NMF. {Based on these assumptions and our derivation in section 3, speech and noise latent variables can be obtained from the observed signal. The whole recognition process is shown in Fig.~\ref{fig:Bayesian_model}(b).}

{To sum up, we intend to first estimate the latent variable distributions of the speech $p(\mathbf{z}_x|\mathbf{y} )$ and the noise $p(\mathbf{z}_d|\mathbf{y} )$ from the observed signal to acquire latent variables $\mathbf{z}_x$ and $\mathbf{z}_d$, respectively. After that, we use the estimated latent variables as the input of the decoder of C-VAE and N-VAE to obtain the probability distribution of $q(\mathbf{x}|\mathbf{z}_x )$ and $q(\mathbf{d}|\mathbf{z}_d )$ for SE.} 

\vspace{-0.4cm}
\section{SE with Bayesian permutation training}
\label{sec:SE_bayesian}
\vspace{-0.2cm}
{\subsection{Variational autoencoder with multiple latent variables}
\vspace{-0.2cm}

VAE \cite{kingma2013auto} defines a probabilistic generative process between observed signal and its latent variables and provides a principled method to jointly learn latent variables, generative and recognition models, which is achieved by maximizing variational lower bound using stochastic gradient descent (SGD) algorithm. This optimizing process \cite{kingma2013auto} is equal to minimize Kullback-Leibler (KL) divergence ($D_{KL}$) between real  joint probability distribution $p(\mathbf{y},\mathbf{z}_x,\mathbf{z}_d)$ and its estimation $q(\mathbf{y},\mathbf{z}_x,\mathbf{z}_d)$. This process can be written as follows:
\begin{equation}
  \scriptsize
\setlength{\abovedisplayskip}{3pt}
 \begin{aligned}
   &  D_{KL}\left({p({{\bf y},{\bf z}_x,{\bf z}_d})}||{q({{\bf y},{\bf z}_x,{\bf z}_d})}\right) = {\mathbb E_{{\bf {y}} \sim p(\bf {y})}} \left[\log {p({\bf {y}})}\right] \\
   & + {\mathbb E_{{\bf {y}} \sim p(\bf {y})}} [ D_{KL}\left({p({\bf z}_x,{\bf z}_d|{\bf{y}}))}||{q({{\bf y},{\bf z}_x,{\bf z}_d})}\right)].
   \end{aligned}
  \label{KL_basic_noisy_1}
\setlength{\belowdisplayskip}{3pt}
\end{equation}
In (\ref{KL_basic_noisy_1}), the term ${\mathbb E_{{\bf {y}} \sim p(\bf {y})}} \left[\log {p({\bf {y}})}\right]$ is a constant, so minimizing their KL divergence is equal to {minimizing}
\begin{equation}
  \scriptsize
 \setlength{\abovedisplayskip}{3pt}
 \begin{aligned}
 & \mathcal{L} (\theta_y, \varphi_y, \theta_x, \varphi_x, \theta_d, \varphi_d; {\bf y}) \\
  & \quad = {\mathbb E_{{\bf {y}} \sim p(\bf {y})}} [ D_{KL}\left({p({\bf z}_x,{\bf z}_d|{\bf{y}}))}||{q({{\bf y},{\bf z}_x,{\bf z}_d})}\right)] \\
  & \quad = {\mathbb E_{{\bf {y}} \sim p(\bf {y})}} \left[D_{KL}\left({p({\bf z}_x,{\bf z}_d|{\bf{y}}))}||{q({{\bf z}_x,{\bf z}_d})}\right)\right]\\
  &  \quad \quad - {\mathbb E_{{\bf {y}} \sim p(\bf {y})}} \left[ {\mathbb E_{{{\bf {z}}_d,{\bf {z}}_x} \sim p({{\bf {z}}_d,{\bf {z}}_x}|\bf {y})}}\left[\log {q({\bf{y}}|{\bf z}_x,{\bf z}_d)} \right]\right],
   \end{aligned}
  \label{basic_loss_funtion}
  \setlength{\belowdisplayskip}{3pt}
\end{equation}
where $\theta_y, \varphi_y, \theta_x, \varphi_x, \theta_d, \varphi_d$ {are} the parameters that are used to conduct the related probability estimation. The details will be presented later. Here, $-\mathcal{L}$ can be seen as the VAE variational lower bound with multiple latent variables (${\mathbb E_{{\bf {y}} \sim p(\bf {y})}[\log q(\bf {y})]} \ge -\mathcal{L}$) \cite{kingma2013auto}. Minimizing $\mathcal{L}$ is equal to maximize this variational lower bound. Based on our assumptions in section \ref{sec:problem} (${\bf z}_x$ and ${\bf z}_d$ are independent and $p({\mathbf{z}_x},{\mathbf{z}_d}|\mathbf{y}) = p(\mathbf{z}_x|\mathbf{y})p(\mathbf{z}_d|\mathbf{y})$), (\ref{basic_loss_funtion}) can be further written as
 \begin{equation}
  \scriptsize
 \setlength{\abovedisplayskip}{3pt}
 \begin{aligned}
  & \mathcal{L} (\theta_y, \varphi_y, \theta_x, \varphi_x, \theta_d, \varphi_d; {\bf y})\\
  & \quad = {\mathbb E_{{\bf {y}} \sim p(\bf {y})}} \left[ D_{KL}\left({p({\bf z}_x|{\bf{y}})}||{q({\bf z}_x)}\right)\right] \\
  & \quad \quad + {\mathbb E_{{\bf {y}} \sim p(\bf {y})}} \left[  D_{KL}\left({p({\bf z}_d|{\bf{y}})}||{q({\bf z}_d)}\right)  \right] \\ 
  & \quad \quad - {\mathbb E_{{\bf {y}} \sim p(\bf {y})}} \left[ {\mathbb E_{{{\bf {z}}_d,{\bf {z}}_x} \sim p({{\bf {z}}_d,{\bf {z}}_x}|\bf {y})}}\left[\log {q({\bf{y}}|{\bf z}_x,{\bf z}_d)} \right]\right],
   \end{aligned}
  \label{basic_loss_funtion_2}
  \setlength{\belowdisplayskip}{3pt}
\end{equation}}

\vspace{-0.5cm}
{\subsection{DRL with Bayesian permutation training}
\vspace{-0.10cm}

{To estimate the speech and noise latent variables from the observed signal using (\ref{basic_loss_funtion_2}), we propose a Bayesian permutation training process between NS-VAE, C-VAE, and N-VAE.} First, the C-VAE and N-VAE are separately pre-trained using the general VAE training method \cite{kingma2013auto} without supervision. The purpose is to acquire the posterior estimates $p({\bf z}_x|{\bf{x}})$ and $p({\bf z}_d|{\bf{d}})$. Then, the NS-VAE is trained with the supervision of C-VAE and N-VAE.

In (\ref{basic_loss_funtion_2}), the calculation of the first and second term is similar, so we will only use the first term to show the Bayesian permutation process. To achieve supervision learning, we add an attention mechanism ($ {p({\bf z}_x|{\bf x})}/{p({\bf z}_x|{\bf x})}$) for the calculation of first term in (\ref{basic_loss_funtion_2}). Thus, its calculation can be written as (\ref{first_loss_funtion_3})
\begin{equation}
  \scriptsize
\setlength{\abovedisplayskip}{1pt}
 \begin{aligned}
  &  {\mathbb E_{{\bf {y}} \sim p(\bf {y})}} \left[ D_{KL}\left({p({\bf z}_x|{\bf{y}})}||{q({\bf z}_x)}\right)\right] \\
  & \quad = {\mathbb E_{{\bf {y}} \sim p(\bf {y})}} \left[\int p({\bf z}_x|{\bf{y}})\log \frac{p({\bf z}_x|{\bf{y}})p({\bf z}_x|{\bf x})}{q({\bf z}_x)p({\bf z}_x|{\bf{x}})} \, d{\bf z}_x \right] \\ 
   & \quad = {\mathbb E_{{\bf {y}} \sim p({\bf {y}}), {\bf {x}} \sim p(\bf {x})}} \left[ D_{KL}\left({p({\bf z}_x|{\bf{y}})}||{p({\bf z}_x|{\bf{x}})}\right) \right] \\
  & \quad \quad + {\mathbb E_{{\bf {y}} \sim p({\bf {y})}}} \left[ {\mathbb E_{{{\bf {z}}_x} \sim p({{\bf {z}}_x}|{\bf {y}})}}[\log \frac{p({\bf z}_x|{\bf x})}{q({\bf z}_x)} ] \right],
   \end{aligned}
  \label{first_loss_funtion_3}
  \setlength{\belowdisplayskip}{3pt}
\end{equation}
\vspace{-0.10cm}
\begin{equation}
  \scriptsize
 \begin{aligned}
 \setlength{\abovedisplayskip}{0pt}
  & \mathcal{L} (\theta_y, \varphi_y, \theta_x, \varphi_x, \theta_d, \varphi_d; {\bf y}) \\
  & \quad = {\mathbb E_{{\bf {y}} \sim p(\bf {y}),{\bf {x}} \sim p(\bf {x})}} \{D_{KL}\left({p({\bf z}_x|{\bf{y}})}||{p({\bf z}_x|{\bf{x}})}\right) \\
  & \quad \quad + {\mathbb E_{{{\bf {z}}_x} \sim p({{\bf {z}}_x}|{\bf {y}})}}[\log \frac{p({\bf z}_x|{\bf x})}{q({\bf z}_x)}]\} \\
  & \quad \quad + {\mathbb E_{{\bf {y}} \sim p({\bf {y}}), {\bf {d}} \sim p({\bf {d}})}} \{D_{KL}\left({p({\bf z}_d|{\bf{y}})}||{p({\bf z}_d|{\bf{d}})}\right) \\
  &  \quad \quad + {\mathbb E_{{{\bf {z}}_d} \sim p({{\bf {z}}_d}|{\bf {y}})}}[\log \frac{p({\bf z}_d|{\bf d})}{q({\bf z}_d)}]\} \\
  & \quad \quad - {\mathbb E_{{\bf {y}} \sim p(\bf {y})}} \left[ {\mathbb E_{{{\bf {z}}_d,{\bf {z}}_x} \sim p({{\bf {z}}_d,{\bf {z}}_x}|\bf {y})}}\left[\log {q({\bf{y}}|{\bf z}_x,{\bf z}_d)} \right]\right].
   \end{aligned}
  \label{final_loss_funtion}
  \setlength{\belowdisplayskip}{2pt}
\end{equation}
In (\ref{first_loss_funtion_3}), we introduce posterior $p({\bf z}_x|{\bf{x}})$ estimated from C-VAE to conduct supervised latent variable learning. The purpose is to obtain speech latent variables from observed signal. Finally, {substituting (\ref{first_loss_funtion_3}) into (\ref{basic_loss_funtion_2}),} the final loss function can be written as (\ref{final_loss_funtion}). In (\ref{final_loss_funtion}), we can find KL divergence constraints for speech and noise latent variables, which ensures that we can estimate the desired posterior distributions from noisy signal in a supervision way. This is also why our method can disentangle latent variables, and the noise can be estimated by nonlinear VAE rather than linear NMF, which is different from the previous VAE-based SE methods \cite{leglaive2018variance,bando2018statistical,leglaive2019semi,carbajal2021guided,fang2021variational}. Moreover, in (\ref{final_loss_funtion}), $-\mathcal{L}$ can be used as a novel variational lower bound to perform supervised VAE training in other VAE-related applications. To better minimize (\ref{final_loss_funtion}), we introduce C-VAE and N-VAE to conduct joint training, which forms a Bayesian permutation training process between the three VAEs. Finally, the NS-VAE's training loss is
\vspace{-0.2cm}
\begin{equation}
\footnotesize
 \begin{aligned}
 \setlength{\abovedisplayskip}{0pt}
  \mathcal{L}_{total} &= \mathcal{L} (\theta_y, \varphi_y, \theta_x, \varphi_x, \theta_d, \varphi_d; {\bf y}) \\
  & \quad + \mathcal{L}_{c} (\theta_x, \varphi_x; {\bf x}) + \mathcal{L}_{n} (\theta_d, \varphi_d; {\bf d}),
   \end{aligned}
  \label{final_pvaeloss_funtion}
  \setlength{\belowdisplayskip}{2pt}
\end{equation}
where $\mathcal{L}_{c} (\theta_x, \varphi_x; {\bf x})$ and $\mathcal{L}_{n} (\theta_d, \varphi_d; {\bf d})$ are the general VAE loss function for speech and noise, which can be written as 
\vspace{-0.2cm}
\begin{equation}
  \scriptsize
 \begin{aligned}
 \setlength{\abovedisplayskip}{0pt}
   \mathcal{L}_{c} (\theta_x, \varphi_x; {\bf x}) &= {\mathbb E_{{\bf {x}} \sim p({\bf {x}})}} \{ D_{KL}\left({p({{\bf {z}}_x}|{\bf{x}})}||{q({\bf z}_x)}\right) \\
  & \quad - {\mathbb E_{{\bf {z}}_x \sim p({{\bf {z}}_x}|{\bf{x}})}} [\log {q({\bf x}|{\bf z}_x)} ]\},
   \end{aligned}
  \label{clean_vae}
  \setlength{\belowdisplayskip}{2pt}
\end{equation}
\begin{equation}
  \scriptsize
 \begin{aligned}
 \setlength{\abovedisplayskip}{0pt}
   \mathcal{L}_{n} (\theta_d, \varphi_d; {\bf d}) &= {\mathbb E_{{\bf {d}} \sim p({\bf {d}})}} \{ D_{KL}\left({p({{\bf {z}}_d}|{\bf{d}})}||{q({\bf z}_d)}\right) \\
  & \quad - {\mathbb E_{{\bf {z}}_d \sim p({{\bf {z}}_d}|{\bf{d}})}} [\log {q({\bf d}|{\bf z}_d)} ]\}.
   \end{aligned}
  \label{noise_vae}
  \setlength{\belowdisplayskip}{2pt}
\end{equation}
In (\ref{final_pvaeloss_funtion}), it can be found that the NS-VAE's training also includes the training of C-VAE and N-VAE, which improves NS-VAE's ability to disentangle latent variables. Minimizing $\mathcal{L}_{total}$ is our final target.

Fig.~\ref{fig:Bayesian_DNN} shows the proposed framework. To summarize, the proposed method includes a training and an enhancement stage. The whole training process can be described as follows: first, C-VAE and N-VAE are separately pre-trained without supervision using (\ref{clean_vae}) and (\ref{noise_vae}). Then, the LPS features of speech, noise and observed signal are separately used as the encoder input of C-VAE, N-VAE, and NS-VAE to estimate posterior distributions $p({{\bf {z}}_x}|{\bf {y}}), p({{\bf {z}}_d}|{\bf {y}}), p({{\bf {z}}_x}|{\bf {x}})$, $ p({{\bf {z}}_d}|{\bf {d}})$ and prior distribution ${q({\bf{y}}|{\bf z}_x,{\bf z}_d)}$, ${q({\bf{x}}|{\bf z}_x})$, ${q({\bf{d}}|{\bf z}_d)}$. Finally, (\ref{final_pvaeloss_funtion}) is used as a loss function to perform related parameters update with the Adam algorithm \cite{kingma2014adam}. The training is completed when the neural networks converge. In the online SE stage, we assume that {the} ${{\bf z}_x}$ sampled from $p({\bf z}_x|{\bf{x}})$ is approximately equal to the sample ${{\bf z}_x}$ sampled from $p({\bf z}_x|{\bf{y}})$. Therefore, we can separately use {the} NS-VAE encoder's two outputs as input of C-VAE and N-VAE to obtain the prior distributions $q({\bf{x}}|{\bf z}_x)$ and $q({\bf{d}}|{\bf z}_d)$. After that, using {the} reparameterization trick and Monte Carlo estimate (MCE)\cite{kingma2013auto}, the speech and noise signal can be obtained. The enhanced speech is acquired by direct waveform reconstruction or post-filtering methods. This enhanced process is shown in Fig.~\ref{fig:Enhance framework} (a).

\begin{figure}[!tbp]
  \centering
  \centerline{\includegraphics[scale=0.45]{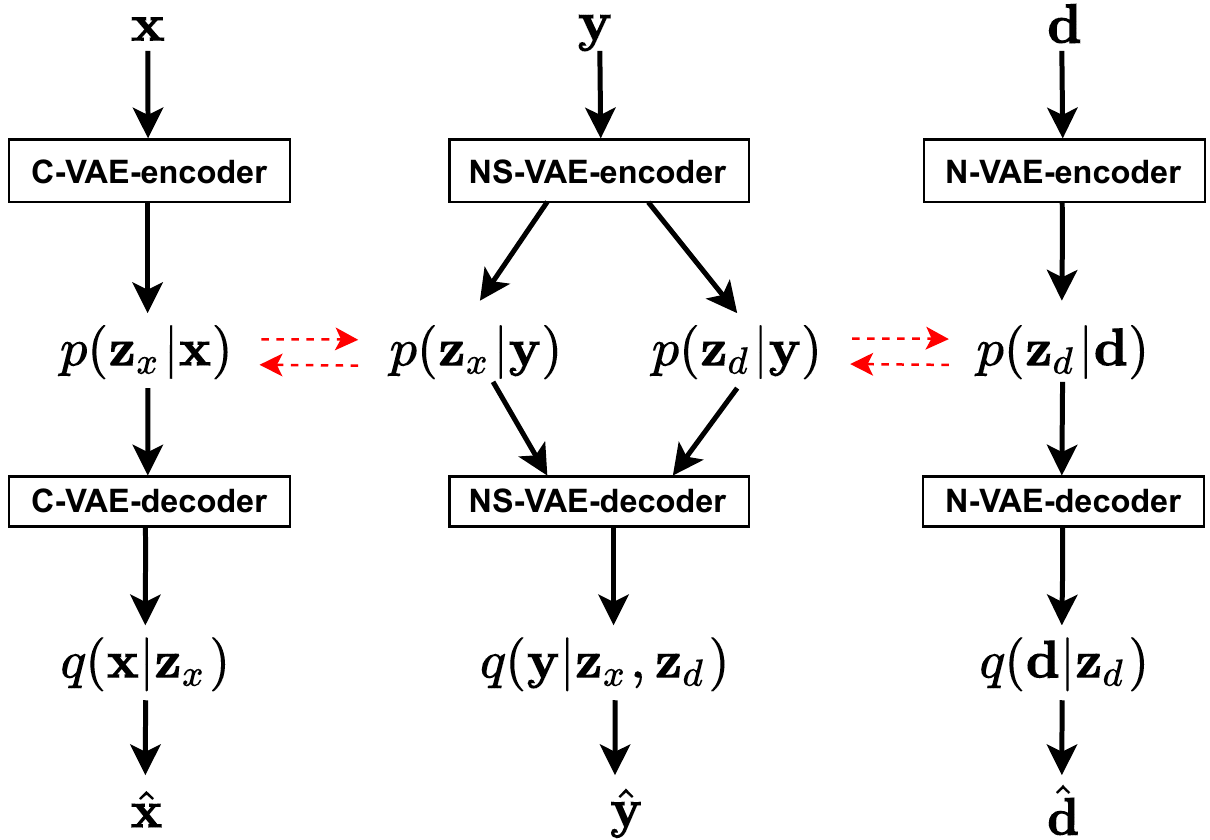}}
  \setlength{\abovecaptionskip}{0.0cm}
  \caption{Model of Bayesian permutation training for SE.}
  \label{fig:Bayesian_DNN}
  \vspace{-0.7cm}
\end{figure}
}
 \vspace{-0.2cm}
\vspace{-0.2cm}
 {\subsection{Calculation of loss function}
\vspace{-0.2cm}
In (\ref{final_pvaeloss_funtion}), the related posterior and prior distributions need to be determined, and $q({{\bf z}_x})$ and $q({{\bf z}_d})$ need to be predefined for the calculation. Here, for the simplicity of calculation, we assume that all the posterior and prior distributions are multivariate normal distributions with diagonal covariance\cite{kingma2013auto}, which is similar to the previous VAE-based SE methods\cite{leglaive2018variance,bando2018statistical,leglaive2019semi,carbajal2021guided,fang2021variational}. For the NS-VAE, we have 
 \begin{equation}
 \scriptsize
  \setlength{\abovedisplayskip}{0pt}
   \begin{aligned}
   & {p({\bf z}_x|{\bf{y}})} = \mathcal{N} \left({{\bf {{\bf z}}}_x;{\mu}_{\theta_{yx}}({\bf{y}}),{\sigma}_{\theta_{yx}}^2({\bf y})\bf{I}}\right), \\
   & {p({\bf z}_d|{\bf{y}})} = \mathcal{N} \left({{\bf {{\bf z}}}_d;{\mu}_{\theta_{yd}}({\bf{y}}),{\sigma}_{\theta_{yd}}^2({\bf y})\bf{I}}\right), \\
   & q({\bf{y}}|{\bf z}_x,{\bf z}_d) = \mathcal{N} \left({{\bf {{\bf y}}};{\mu}_{\varphi_{y}}({{\bf z}_x,{\bf z}_d}),{\sigma}_{\varphi_{y}}^2({{\bf z}_x,{\bf z}_d})\bf{I}}\right),
  \end{aligned}
  \label{noisy_prior}
  \setlength{\belowdisplayskip}{2pt}
\end{equation}
where $\bf I$ is the identity matrix. ${\mu}_{\theta_{yx}}({\bf{y}}),{\sigma}_{\theta_{yx}}^2({\bf{y}}), {\mu}_{\theta_{yd}}({\bf{y}})$, ${\sigma}_{\theta_{yd}}^2({\bf y})$ can be estimated by NS-VAE's encoder $G_{\theta_y}({\bf{y}})$ with parameter ${\theta_y}=\{\theta_{yx}, \theta_{yd}\} $, and ${\sigma}_{\varphi_{y}}^2({{\bf z}_x,{\bf z}_d})$ and ${\mu}_{\varphi_{y}}({{\bf z}_x,{\bf z}_d})$ can be estimated by NS-VAE's decoder $G_{\varphi_y}({{\bf z}_x,{\bf z}_d})$ with parameter ${\varphi_y}$. Due to the space limitation and the fact that the frameworks of C-VAE and N-VAE are similar, we only give the details of C-VAE. Here, we have
\begin{equation}
 \scriptsize
  \setlength{\abovedisplayskip}{0pt}
   \begin{aligned}
   & {p({\bf z}_x|{\bf{x}})} = \mathcal{N} \left({{\bf {{\bf z}}}_x;{\mu}_{\theta_{x}}({\bf{x}}),{\sigma}_{\theta_{x}}^2({\bf x})\bf{I}}\right) \\
   & q({\bf{x}}|{\bf z}_x) = \mathcal{N} \left({{\bf {{\bf x}}};{\mu}_{\varphi_{x}}({{\bf z}_x}),{\sigma}_{\varphi_{x}}^2({{\bf z}_x})\bf{I}}\right),
  \end{aligned}
  \label{posterior}
  \setlength{\belowdisplayskip}{2pt}
\end{equation}
where ${\mu}_{\theta_{x}}({\bf{x}}), {\sigma}_{\theta_{x}}^2({\bf x})$ are obtained by C-VAE's encoder $G_{\theta_x}({\bf{x}})$ with parameter ${\theta_x}$, and ${\mu}_{\varphi_{x}}({{\bf z}_x})$, ${\sigma}_{\varphi_{x}}^2({{\bf z}_x})$ can be estimated by C-VAE's decoder $G_{\varphi_x}({{\bf z}_x})$ with parameter ${\varphi_x}$. $q({{\bf {z}}_d})$ and $q({{\bf {z}}_x})$ are pre-defined as a centered isotropic multivariate Gaussian  $q({\bf z}_x) = \mathcal{N} ({{\bf {{\bf z}}}_x;{\bf{0}},{\bf{I}}})$ and $q({\bf z}_d) = \mathcal{N} ({{\bf {{\bf z}}}_d;{\bf{0}},{\bf{I}}})$. Finally, when all the distributions are determined, we can apply loss function (\ref{final_pvaeloss_funtion}) and the Adam algorithm to perform related parameters update for SE. 
} 

\begin{figure}[!tbp]
  \centering
  \centerline{\includegraphics[scale=0.45]{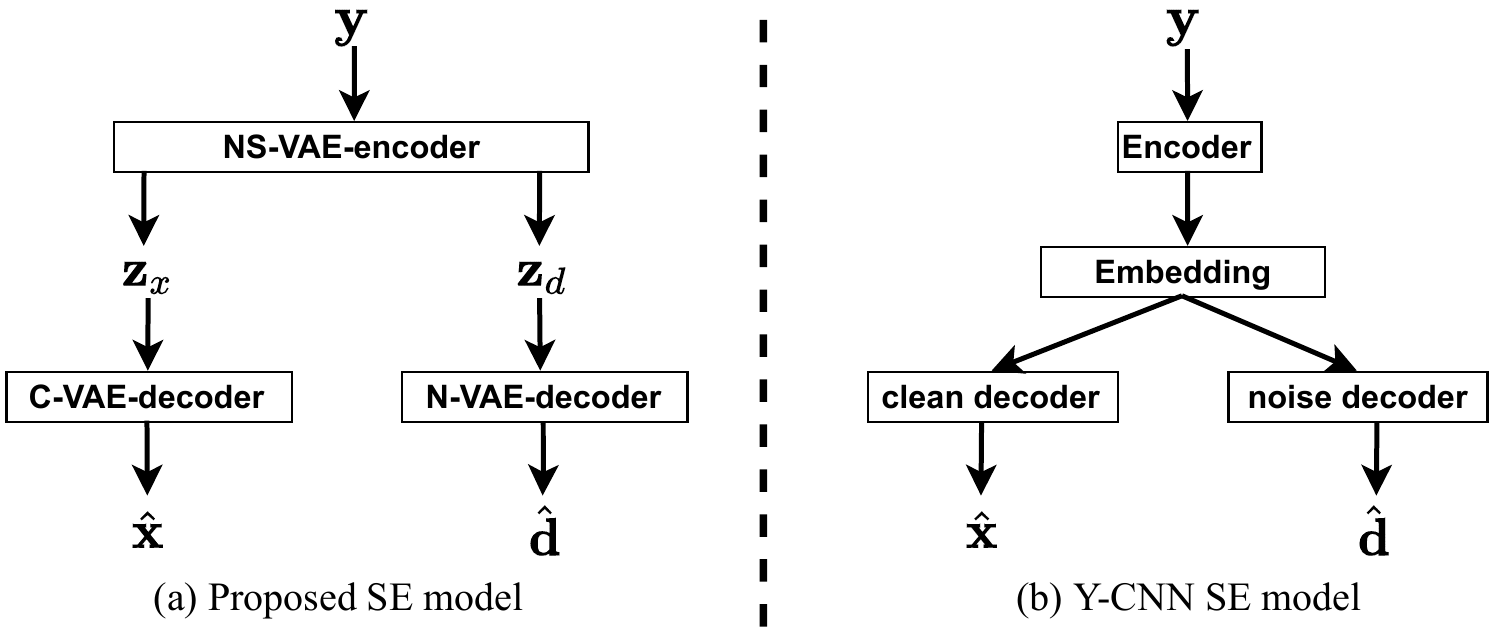}}
  \setlength{\abovecaptionskip}{0.2cm}
  \caption{Enhancement framework comparison.}
  \label{fig:Enhance framework}
  \vspace{-0.3cm}
\end{figure}

\begin{figure}[!tbp]
  \centering
  \centerline{\includegraphics[scale=0.45]{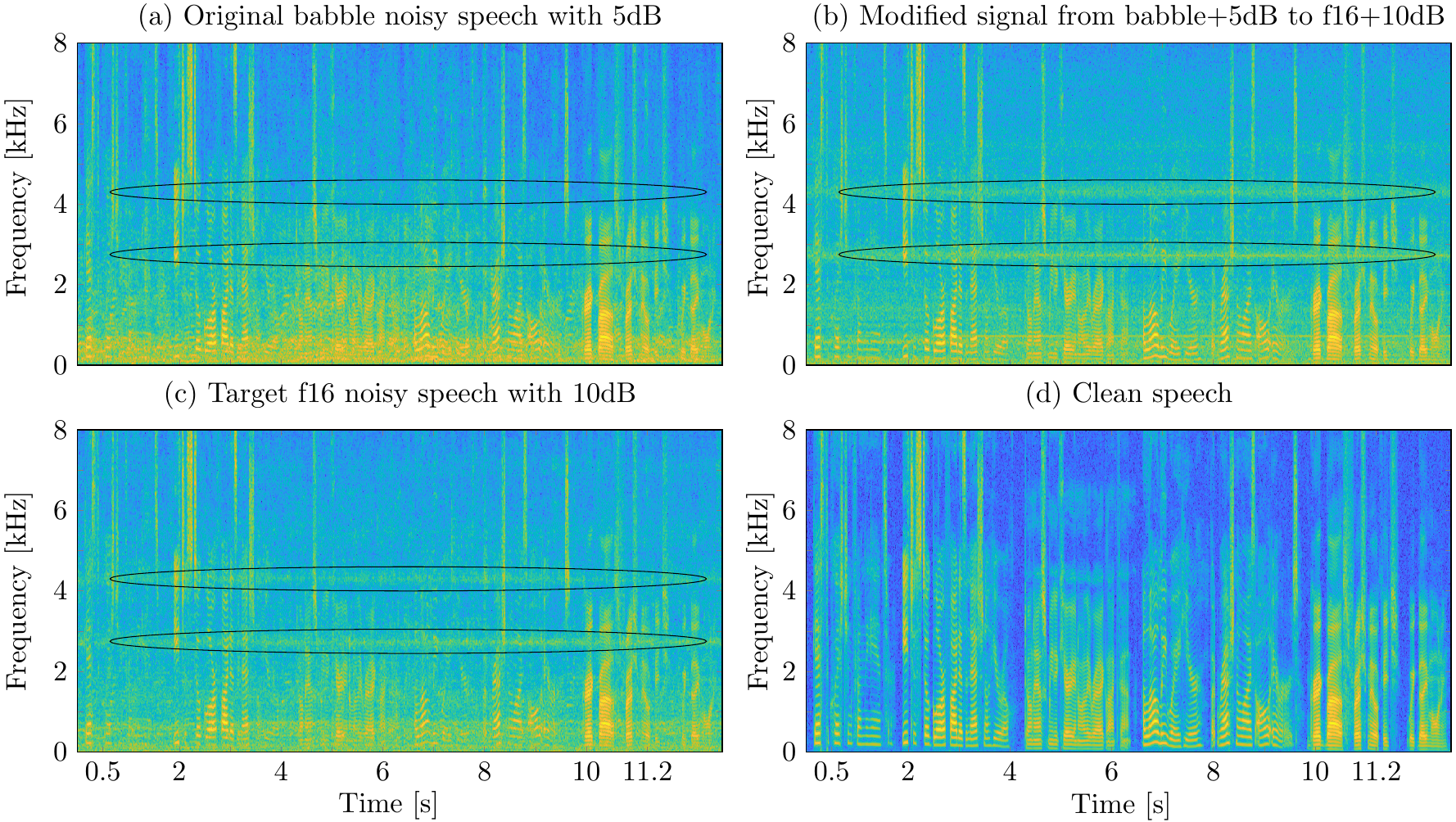}}
  \setlength{\abovecaptionskip}{0.1cm}
  \caption{{Experiment for disentangling latent variables.}}
  \label{fig:spec}
  \vspace{-0.6cm}
\end{figure}
\vspace{-0.35cm}
\section{Experiment and result analysis}
\vspace{-0.3cm}
\label{sec:experiment}
{In this section, the proposed algorithm is evaluated. At first, we will use an example to verify that the proposed method can disentangle different latent variables from the observed signal. After that, an experiment will show the SE performance of our method.}

{\bf Dataset:} In this work, we use the TIMIT database\cite{garofolo1993darpa}, 100 environmental noises\cite{hu2010tandem}, DEMAND database\cite{thiemann2013diverse} and NOISEX-92 database\cite{varga1993assessment} to evaluate the performance of the proposed algorithm. In the training stage, the Babble, F16 noise from the NOISEX-92 database\cite{varga1993assessment} and 90 environmental noise (N1--N90)\cite{hu2010tandem} are used to conduct experiments. All 4620 utterances from the TIMIT database are corrupted by 92 types of noise at four different signal-to-noise ratio (SNR) levels, {i.e., -5, 0, 5, and 10 dB.} The utterances are randomly selected from these corrupted utterances, and they are connected to a 12-hour noisy speech database. Meanwhile, the corresponding speech and noise databases are also obtained. In the test stage, 200 utterances from the TIMIT test set, including 1680 utterances, are randomly chosen to build the test database. 13 types of noise (office\cite{thiemann2013diverse}, factory\cite{varga1993assessment}, and 10 unseen environmental noise (N91—N100)\cite{hu2010tandem}) are randomly added to the 200 utterances at four SNR levels (i.e., -5, 0, 5, and 10 dB) to conduct experiment. In our experiments, all the signals are down-sampled to 16 kHz. The frame length is 512 samples with a frame shift of 256 samples. 
\vspace{-1pt}

{\bf Baseline:} To evaluate the performance of the proposed method, we use a supervised SE model as a reference method (referred to Y-CNN) \cite{huang2014deep}. This is similar to the proposed method and can perform SE by {direct waveform reconstruction}\cite{xu2014regression} or estimated mask \cite{wang2014training}. For a fair comparison, we use a convolutional neural network (CNN) to replace the original DNN\cite{huang2014deep} to improve its performance. Fig.~\ref{fig:Enhance framework} shows the framework comparison of enhancement. Y-CNN has the same encoder and decoder as the proposed model. The only difference between Y-CNN and our method is the training loss function. The loss function of the proposed method applies deep representation learning (DRL) and reasonable assumptions to disentangle latent variables, which is not achieved by Y-CNN \cite{huang2014deep}.

{\bf Experimental setups:} There are three VAEs in our proposed method. The C-VAE and N-VAE have the same structure. 1D CNN which is widely used in SE \cite{ luo2019conv} is adopted in the experiment. C-VAE' encoder includes four hidden 1D convolutional layers. The number of channels in each layer is 32, 64, 128, and 256. The size of each convolving kernel is 3. The two output layers of the encoders are fully connected layers with 128 nodes. By using the reparameterization trick, the decoders' input size can also be set as 128. The decoder consists of four hidden 1D convolutional  layers (the channel number of each layer is 256, 128, 64, and 32 with 3 kernel) and two fully connected output layers with 257 nodes. Moreover, the activation functions for the hidden and output layer are ReLU and linear activation function, respectively. For NS-VAE, its encoder also includes four 1D convolutional layers with ReLU as the activation function. The other parameter setting is the same as C-VAE. Additionally, its encoder has four output layers with 128 nodes and a linear activation function. The input size of the NS-VAE decoder is 256, which includes the latent speech and the noise variables. The decoder structure of NS-VAE's decoder is the same as that of C-VAE (Y-CNN's encoder and decoders only have one output layer with  128 and 257 nodes, respectively, because it does not disentangle latent variables. The other settings are the same as that of NS-VAE). In the training stage, all networks are trained by the Adam algorithm with a 128 mini-batch size. The learning rate is 0.001.
\begin{table*}[!t]

\setlength{\tabcolsep}{2mm} 
 \centering
 \setlength{\abovecaptionskip}{0.05cm}
  \caption{{\scriptsize{Average PESQ and SI-SDR comparison of different methods}}}
  \label{tab: score_comparison}
  \centering
   \tiny
    \begin{tabular}{ccccccccccc}
    \toprule
    \multirow{2}*{SNR}& \multicolumn{5}{c}{SI-SDR (dB)} &\multicolumn{5}{c}{PESQ} \\
    \cmidrule(lr){2-6} \cmidrule(lr){7-11}
    &Noisy &Y-L &PVAE-L &Y-M &PVAE-M& Noisy &Y-L & PVAE-L &Y-M &PVAE-M\\
    \midrule
    -5 &-5.67($\pm\,$0.22)&1.25($\pm\,$0.67) & 2.84($\pm\,$0.72) &2.04($\pm\,$0.68) & {\bf4.01($\pm\,$0.88)}& 1.43($\pm\,$0.02) &1.59($\pm\,$0.03)&{\bf1.87($\pm\,$0.03)} &1.68($\pm\,$0.03) &{1.86($\pm\,$0.03)} \\
    \cmidrule(lr){1-11}
    0 &-0.69($\pm\,$0.22) &4.52($\pm\,$0.47) &6.32($\pm\,$0.48) &7.40 ($\pm\,$0.68) &{\bf8.59($\pm\,$0.75)} & 1.78($\pm\,$0.02) &2.02($\pm\,$0.02) &2.24($\pm\,$0.03) &2.11($\pm\,$0.03) &{\bf2.27($\pm\,$0.03)}\\
    \cmidrule(lr){1-11}
    5 &4.30($\pm\,$0.23)&6.76($\pm\,$0.29) &8.67($\pm\,$0.31) &11.74($\pm\,$0.62) &{\bf12.33($\pm\,$0.61)} &2.13($\pm\,$0.02) &2.43($\pm\,$0.02) &2.57($\pm\,$0.02) &2.53($\pm\,$0.03) &{\bf2.63($\pm\,$0.02)} \\
    \cmidrule(lr){1-11}
    10 &7.30($\pm\,$0.23) &8.05($\pm\,$0.18)&10.03($\pm\,$0.23) &15.17($\pm\,$0.54) &{\bf15.41($\pm\,$0.50)} & 2.46($\pm\,$0.01) &2.76($\pm\,$0.02) &2.80($\pm\,$0.02) &2.86($\pm\,$0.02) &{\bf2.91($\pm\,$0.02)} \\
    \bottomrule                             
  \end{tabular}
    
    \vspace{-0.5cm}
\end{table*}

{{\bf Experimental results:} Firstly, we will verify the ability of the proposed method to effectively disentangle the speech and noise latent variables from observed signals. Based on our assumption, the observed signal $\mathbf{y}$ is determined by $\mathbf{z}_x$ and $\mathbf{z}_d$. Thus, if we use different $\mathbf{z}_x$ or $\mathbf{z}_d$ as the input of NS-VAE's decoder, we can obtain the different observed signal. $\mathbf{z}_x$ and $\mathbf{z}_d$ can be acquired by different NS-VAE encoders. Fig.~\ref{fig:spec} shows the experimental result. In this example, we first disentangle the latent variables of the observed signal (babble noise with 5dB). Then, we keep the speech latent variable $\mathbf{z}_x$ and replace the noise latent variable with another noise latent variable (f16 noise with 10dB). Finally, the new combination of latent variables is used as the input of NS-VAE's decoder to acquire the modified signal. Fig.~\ref{fig:spec}(b) shows the modified signal. Comparing Fig.~\ref{fig:spec}(a), (b), (c), and (d), it can be found that the modified signal successfully removes the babble noise character, and the original noise character is replaced by f16 noise character. (The modified signal has the constant noise around 3000 and 4000 Hz as shown in the black circle area, which is the same as the target signal.) Furthermore, the modified signal also preserves the original speech character. Therefore, this experiment indicates that the proposed method can effectively disentangle different latent variables.
\begin{table}[!t]
\setlength{\tabcolsep}{2mm} 
 \centering
 \setlength{\abovecaptionskip}{0.05cm}
  \caption{{\scriptsize{Average STOI comparison of different methods}}}
  \label{tab: STOI_score_comparison}
  \centering
   \tiny
    \begin{tabular}{cccccc}
    \toprule
    SNR &Noisy &Y-L &PVAE-L &Y-M &PVAE-M\\
    \midrule
    -5 &57.62($\pm\,$1.31)&57.63($\pm\,$1.67) &60.00($\pm\,$1.33) &59.72($\pm\,$1.70) &60.32($\pm\,$1.40)\\
    \cmidrule(lr){1-6}
    0 &70.02($\pm\,$1.24) &69.80($\pm\,$1.48) &70.68($\pm\,$1.12) &72.02($\pm\,$1.43) &71.75($\pm\,$1.19)\\
    \cmidrule(lr){1-6}
    5 &80.20($\pm\,$0.90)&79.20($\pm\,$1.18) &79.87($\pm\,$0.87) &81.96($\pm\,$1.02) &80.78($\pm\,$0.91)\\
    \cmidrule(lr){1-6}
    10 &86.32($\pm\,$0.50) &85.60 ($\pm\,$0.72)&84.32($\pm\,$0.54) &88.80($\pm\,$0.63) &87.24($\pm\,$0.58)\\
    \bottomrule                             
  \end{tabular}
    \vspace{-0.7cm}
\end{table}

In the second experiment, all algorithms are evaluated by the scale-invariant signal-to-distortion ratio (SI-SDR) in decibel (dB) \cite{le2019sdr}, short-time objective intelligibility (STOI)\cite{taal2011algorithm}, and perceptual evaluation of speech quality (PESQ)\cite{rix2001perceptual}. The enhanced speech is obtained by direct waveform reconstruction \cite{xu2014regression} or  soft time-frequency mask estimation \cite{huang2014deep}. We use Y-M and Y-L to represent that the enhanced speech is acquired by mask estimation and direct waveform reconstruction using Y-CNN, respectively. Similarly, PVAE-L and PVAE-M denote that the enhanced speech is obtained by the proposed method using direct waveform reconstruction and mask estimation, respectively. {Table.~\ref{tab: score_comparison} and \ref{tab: STOI_score_comparison} show the PESQ, SI-SDR, and STOI comparisons with  a 95\%  confidence  interval.} The results verify that our method can learn latent speech and noise variables from observed signals because PVAE-L achieves better PESQ and SI-SDR performance than Y-L. This means that C-VAE's decoder can recognize speech latent variables that are disentangled by NS-VAE. Moreover, PVAE-M significantly achieves better PESQ and SI-SDR performance than Y-M, which shows that our method can estimate a more accurate mask for SE. This result also illustrates that our approach has better noise estimation performance than the reference method. Additionally, the results also show that Y-CNN's performance can be improved by the proposed loss function. Table.~\ref{tab: STOI_score_comparison} shows that the STOI score is competitive between the proposed and the reference algorithms. We think that the STOI score of the proposed method can be further improved by improving PVAE's disentangling performance \cite{burgess2018understanding}. Overall, PVAE-M achieves the best SE performance across the three evaluation criteria. Here, we only use a basic neural network to verify our algorithm. Its performance can be further improved by using more advanced neural networks and other speech features\cite{wang2018supervised,luo2019conv,li2021icassp}.}
\vspace{-10pt}

\section{Conclusion}
\vspace{-8pt}
{
In this paper, a supervised Bayesian permutation training DRL method is proposed to disentangle latent speech and noise variables from the observed signal for SE. The proposed method is based on VAE and Bayesian theory. The experimental results show that our method cannot only successfully disentangle different latent variables but also obtain higher SI-SDR and PESQ scores than the state-of-the-art reference method. Moreover, the results also illustrate that the SE performance of the reference method can be improved by introducing the proposed DRL algorithm. In future work, some other strategies can be considered to further improve the disentangling performance of latent variables. In addition, the proposed method can also be applied in other speech generative {tasks, e.g., voice conversion and ASR.}}

\vspace{-9pt}
  
\ninept 
\bibliographystyle{IEEEtran}
\bibliography{IEEEabrv,myabrv_new,my_reference}
\end{document}